\documentclass[12pt,preprint]{aastex}
\begin{document}

\title{An assessment of the rotation rates of the host stars of extrasolar planets}

\author{Sydney A. Barnes}
\affil{Department of Astronomy, Univ. of Wisconsin - Madison, WI, USA}

\begin{abstract}

The rotation periods of the host stars of extrasolar planets have been
assessed against those of the Mt. Wilson stars, open cluster stars,
and evolutionary stellar models that include rotation. They appear to 
be normal, modulo certain inconsistencies in various necessary inputs. 
Selection of candidate planet hosts for radial velocity surveys by low
rotation or activity has resulted in a planet host sample skewed towards
older stars. Thus, cross-comparisons must be age dependent. However,
self-consistent ages are difficult to obtain, and activity ages show
signs of systematic errors. There are indications that activity ages
ought to be increased for lower-than-solar-mass stars and decreased for
higher-than-solar-mass stars. Age uncertainties and a scarcity of measured 
rotation periods for planet host stars inflate the dispersion in older stars 
relative to those in open clusters.
The presently available rotational models display inadequacies, most
notably in producing fast-enough early-type stars. 
The fact that only one planet host star, $\tau Boo$, strongly suggests
tidal spin-up, while of order ten systems suggest orbital circularization
is explicable in terms of the differing timescales for these two phenomena.
The rotational normalcy of the planet host stars and other considerations
suggest that they are not especially different from other main sequence
stars and that circumstellar matter and/or planets are probably 
ubiquitous, at least among sufficiently metal-rich solar-type stars.

\end{abstract}


	\section{Introduction}

Over fifty systems are now known to harbor extra-solar planets. The genesis 
of these systems is of obvious interest, but so little is currently known 
about their properties that almost any additional information is useful.
Although the planetary systems discovered thus far, primarily by the 
radial-velocity technique,  differ from the Solar System, the central stars 
themselves do not appear to be particularly distinguished. 
For the most part, they are just `normal' main sequence stars, except that
evidence seems to be accumulating that they are metal-rich relative to
nearby field stars (eg. Butler et al. 2000; Gonzalez et al. 2001; 
Santos et al. 2000).
A large and important question concerning these systems is simply whether
they possess any other defining characteristics.

A smaller, more specific, question is the one suggested by the title: 
Is there anything unusual about their rotation rates?
Many of these systems contain solar-type stars which, for geographical reasons,
have been studied extensively, particularly with reference to their magnetic
and rotational properties. The exemplar of such work is the Mount Wilson HK 
project reviewed in this context by Baliunas et al. (1998).
Also, over the past decades, both theoretical and observational infrastructure 
has been built and developed to understand the rotational evolution of stars. 
In addition to the physics included in standard stellar evolution, such models 
implement rotation in some way, begin nowadays on the pre-main sequence, at 
the stellar birthline, and then evolve onto, on, and beyond the main sequence.
The key early stages are directly comparable to observations of stellar 
rotation rates among TTauri stars and those in young open clusters, enabling 
us to select a preferred set of models which agree with the known observations 
at early ages. Thus, we are in a position to examine whether there is anything 
unusual about the rotation rates of planet host (henceforth PH) stars.

This preferred set of models of the rotational evolution of solar-type stars 
includes magnetic saturation and more importantly in the present case, 
star-disk interaction in the pre-main sequence phase. This latter interaction 
has been suggested as being responsible for the presence of slow rotators in 
young star clusters. This picture would lead us to expect planet host stars to 
be slow rotators at their respective ages. This can now be checked, at least 
in a preliminary way, and that is a subject of this paper. We also use this
opportunity to extend rotational considerations to single stars and older ones 
than those in young open clusters, as exemplified by the Mt. Wilson and PH 
star samples.

We begin by presenting the period and age information available for PH stars,
which we then compare with the Mount Wilson data. Thereafter, we include  
the open cluster data, go on to discuss the models, which we then compare
with the data. After commenting briefly on tidal effects, we present the 
conclusions, and then end with a discussion of some broader issues.

This investigation was originally undertaken with the premise that there 
should be
something special about the rotation rates of PH stars, and indeed there are 
inconsistencies of various kinds, which we discuss herein. On balance, 
however, the information currently available does not suggest that planet host 
stars have unusual rotational properties. This suggests either that planets 
are ubiquitous, at least among solar-type stars, or else that planetary 
companions do not affect their rotation at all.

	\section{Planet Host Star Data}

Assuming that enough basic information is known about the PH star 
(eg. evolutionary state, position on H-R diagram etc.) to ensure that it is
a `normal' main sequence solar-type star (say $0.6-1.2$ $M_{\odot}$), we need 
two additional items in order to compare it with models of stellar angular 
momentum evolution - the star's rotation period\footnote{The $\sin i$ ambiguity in $v \sin i$ observations reduce their usefulness in this context.} and its 
age. The latter, or at least an approximation thereof, is needed because 
stellar rotation rates, even among solar-type stars, change steadily with age 
(cf. Kraft 1967) and cannot all be lumped together. 
Also, the presently available PH star sample is not unbiased with respect to
rotation because the original sample was stripped of fast-rotating or 
high-activity stars. Thus, a comparison to another sample must be 
age dependent. Fortunately, both of these quantities are available, albeit 
with varying quality, for a sizable fraction of these stars.

	\subsection{Rotation periods of PH stars}
 
For eight solar-type stars, as listed in Table 1, and an additional three 
higher-mass stars
listed in Table 2, directly measured rotation periods are available, derived 
through photometric or spectroscopic variability, almost all of them from the 
Mount Wilson Ca\,II HK survey. The early compilations by Vaughan et al. (1981)
and Noyes et al. (1984) have been updated by Baliunas, Sokoloff \& Soon (1996)
and re-assessed in individual cases by Henry et al. (2000) and others as noted
in Tables 1 \& 2. Half of these periods are uncertain (marked by colons in the
tables and the figures).

For those with $B-V$ colors between 0.4 and 1.4 but no directly measured 
periods, the strength of the Ca\,II emission may be used to estimate the 
rotation period via the empirical relationship given by Noyes et al. (1984). 
While these are obviously less secure, the relationship is still invaluable
and indeed Noyes et al. (1984) demonstrated a remarkable agreement between 
the observed and calculated periods (see Table 1 in their paper).
 
We have compiled host star rotation periods from the literature, reproduced 
them in Table 1 and plotted them in Fig.1 and subsequent figures, maintaining 
the important distinction between measured and calculated periods, represented 
using large and small asterisks respectively. Higher mass PH stars (not 
expected to be spun down similarly because of the lack of a surface convection
zone and not modeled here) are also displayed in the 1.2 $M_{\odot}$ bin using 
`x'es. The calculated periods listed 
have mostly been calculated using the Noyes et al. (1984) relation, based
on the $R'_{HK}$ flux (see references to Table 1). However, the $R'_{HK}$ flux
measurements are a heterogeneous dataset, and it is difficult to assess whether
they are all on the same system as the Mount Wilson data. Although we are 
encouraged by the fact that the measured and calculated periods, as displayed 
in Table 1 in Noyes et al. (1984), are similar, it seems important to retain 
this distinction for the 
present, and treat the calculated periods with appropriate caution\footnote{Also note that the activity level of an individual star (including the Sun) can vary over time, influencing a period calculated using only one or two measurements.}.

	\subsection{Ages of PH stars}

For a star cluster, we have a preferred way of deriving the age of the system. 
This is the isochrone scheme in common use today, named as such and applied 
to NGC 188 using modern stellar models by Demarque \& Larson (1964).
For single stars in general, ages are difficult to derive, have large errors, 
and those for the PH stars in particular are plagued by the non-uniformity 
inherent in the use of a heterogeneous set of age determination techniques. 
Fortunately, we are able to tolerate reasonably large uncertainties in stellar 
ages as the surface rotation rate, $v$, of a solar-type star on the main 
sequence, varies only slowly with age, $t$: $v \propto t^{-1/2}$ 
(Skumanich 1972). 

We have obtained ages for all of the stars in Tables 1 and 2 from the 
literature. These are of two kinds - isochrone and activity - and we have
listed both where available. A large fraction of the isochrone ages listed
have been derived by Gonzalez and collaborators from the Schaller et al. 
(1992) and Schaerer et al. (1993) stellar evolutionary sequences. 
The activity age comes from the relationship between chromospheric 
activity, as revealed by various indicators of the Ca\,II HK flux, and the age 
(Wilson 1963; Soderblom, Duncan \& Johnson 1991; Donahue 1998). The recent
practice, also followed here, has been to calculate these from the $R'_{HK}$ 
values, using the relationship from Donahue (1998), as discussed subsequently.

Fig. 2 compares the isochrone and activity ages for the stars where both are
available. It demonstrates that although the agreement between the two 
age-determination techniques is not perfect, they are mostly consistent,
and in reasonable agreement shortward of 8 Gyr. The chromospheric ages in
this range are marginally higher than the isochrone ages with a typical
disagreement of 0.1 dex, which is probably not meaningful in Fig. 1 and
subsequent figures, given the systematic errors that are undoubtedly present.
Beyond 8 Gyr, the chromospheric ages are typically smaller by upto 0.3 dex.
These stars are on the fringes of the distribution, and as will be shown
subsequently, do not contribute significantly to the rotational discussion,
which is not developed sufficiently for evolved stars. Furthermore, the 
deviations could be damped somewhat by combining the different ages.

Lachaume et al. (1999) provide an inter-comparison and assessment of different
age-determination techniques for a group of nearby main sequence stars, six
of which are included in our sample, and have self-consistent ages.
The one truly anomalous case is the F7V star HD 89744, with isochrone
and activity ages of 1.8 and 8.4 Gyr respectively. This star is beyond the
mass range of our models.
For plotting purposes in this paper, the isochrone ages will be 
accorded primacy, and the activity ages used when the former are unavailable.

Although we have used the information at hand,
we are unconvinced of the advisability of deriving the age of a given star
merely by using any readily available instrumental setup to measure an 
$R'_{HK}$ value. The $R'_{HK}$ values ought to be uniformly measured and placed
on the same system as that used for the age calibration, whether it be the 
Mount Wilson or other dataset (see discussion in Soderblom et al. 1991). Thus, 
it might we worthwhile to observe all the PH stars in a uniform manner, as 
undertaken for example by Henry et al. (1996) for Southern solar-type stars, 
and also to include other well-studied comparison stars. 
On a positive note, although the presently available dataset 
is somewhat sparse, it is expanding rapidly, and we can look forward to much 
more progress as additional planetary systems are discovered. 
To mitigate the confusion inherent in a fast-growing field, we have documented 
the source of each number in Tables 1 and 2 and indicated alternate values,
where appropriate, in the notes. We will see subsequently that regardless of
the manner of deriving ages, and unlike a sample of nearby field dwarfs such
as the Mt. Wilson stars, young stars are almost entirely absent from the 
present PH star sample.

	\section{Mount Wilson stars}

The Mount Wilson Ca\,II HK program begun by Olin Wilson is the most 
comprehensive and longest-term study of nearby solar-type stars. It consists 
of a (still-growing) database of observations of 112 stars that have been 
monitored for activity for about 30 years. A great deal is known about these 
stars (Wilson 1963; Wilson \& Skumanich 1964; Wilson \& Wooley 1970; 
Wilson 1978; Duncan et al. 1991; Baliunas et al. 1995; Baliunas et al. 1998).
Furthermore, there is significant overlap between the PH star and Mount Wilson
samples, a fact which naturally suggests an inter-comparison between them.
We know that the Mount Wilson stars are `normal'; perhaps they have partially 
defined what it means to be a `normal' solar-type main sequence star. 
Thus, such a comparison would really tell us whether planet host stars
are different from normal main sequence stars.

We need to know the colors (or masses), the periods, and ages of these stars.
The most recent compilation of the colors and periods, which we adopt for this
study, is that of Baliunas, Sokoloff \& Soon (1996).
In addition, we would like to have an indication of their ages, but it is
more difficult to gauge the ages of single stars outside clusters (as discussed
earlier) or binary star systems, where the stellar tracks that match the known 
properties of stars, usually the luminosity and the temperature at a common 
age, can be used to determine it. See for instance, the work on the visual
binary system $\alpha$CenAB by Guenther \& Demarque (2000).

	\subsection{Comparison with the Vaughan (1980) classification:}

In the case of the Mount Wilson stars, we have a crude classification of the 
sample into Young (Y) and Old (O) stars, by Vaughan (1980) and supplemented by 
Noyes et al. (1984), based on the value of the S flux index and the form of 
the observed cyclical variations. We have chosen the boundary between the 
Young and Old stars to be 2 Gyr, with the Young star bin displaying ones as 
young as 300Myr and the Old star bin displaying those as old as 10 Gyr. These 
divisions are consistent with other available information about these stars, 
but obviously should not be treated as gospel truth. Using the longest and 
shortest observed rotation periods to set the y-boundaries for each bin,
we have represented the Mount Wilson data using light grey boxes in the
upper panels of Fig.1. Although we might expect the Mount Wilson stars to be 
concentrated along the diagonals of these boxes rather than to be peppered 
uniformly across them, nevertheless, a crude comparison seems warranted, 
and so the PH stars are also plotted in Fig.1 using asterisks as mentioned
earlier, and `x'es for higher mass stars. The large square symbol represents 
the Sun.

The first item to note is that the solar- and higher-mass PH stars are mostly
in the old star category. We interpret this to be a consequence of selecting
planet search stars for low activity and rotation, leading to a preferential
selection of old stars. This is even more apparent upon glancing ahead to
the lower panels of Fig.1 which show the distribution by age and rotation of
the Mt. Wilson star sample, one which is wider in both respects.

This comparison demonstrates, to the admittedly crude level of age
discrimination here, that there is no discernable difference between 
the Mount Wilson periods and the PH star periods. There is no clustering
that is immediately obvious, at least not with the present sample size,
and no reason to suspect that these two populations differ in their rotational 
properties, once the ages have been taken into account. 
The PH stars outside or on the edge of the Mt. Wilson boxes
have uncertain properties, and three of them, HD95128, HD 177830, and 
HD 117176 are known to be evolved. We now consider the consequences of 
improving the age resolution of the Mt. Wilson stars.

	\subsection{Comparison with ages as per Soderblom et al. (1991)}

Soderblom, Duncan \& Johnson (1991) have published a relationship\footnote{ 
Actually, they have supplied three relationships, of which we have chosen the third, it being the most comprehensive. The first, although it suggests younger (and perhaps more plausible) ages, does not include their F star data, while the second does not include the solar, Hyades and Ursa Major information.}
between $R'_{HK}$ and age, based on work in binary star systems:
\begin{equation}
 log\,t = (-1.50 \pm 0.03)\,log\,R'_{HK} + (2.25 \pm 0.12)
\end{equation}
Using this relationship, it is possible to assign ages to the Mount Wilson
stars, and thus to compare them in more detail to the PH star data. 
This comparison is displayed in the lower panels of Fig.1 using $+$ signs
to represent the Mount Wilson stars.

As expected, the bins, especially the low-mass one, are `diagonalized'
somewhat by this age assignment. Use of the second relationship from 
Soderblom, Duncan \& Johnson (1991), abbreviated as SDJ91, instead of the 
third does not change this figure appreciably. 
Again, it is clear that to within the errors that doubtlessly are incorporated
in the age and period assignments, there is no difference between the PH stars 
and the Mount Wilson stars. The only PH stars deviating from the Mt. Wilson
trends are those with uncertain periods, as discussed above.
The overall trend of the Mt. Wilson data, also followed by the PH stars, is
that higher mass stars rotate faster.

	\subsection{Comparison with ages as per Donahue (1998)}

A more serious problem with the SDJ91 relationship, pointed out by Donahue 
(1998) is that it substantially and systematically overestimates the ages of 
stars younger than 1 Gyr because it fails to account for the `saturation' in 
activity for young stars. Thus,
Donahue (1998) has modified the SDJ91 relationship to account for young stars
where the SDJ91 relationship seems to fail:
\begin{equation}
 log\,Age = 10.725 - 1.334\,R_5 + 0.4085\,R_5^2 - 0.0522\,R_5^3
\end{equation}
where $ R_5 = 10^5\,R'_{HK} $. For stars older than 1 Gyr,
the two relationships are essentially the same. We have used this modified 
relationship to re-calculate
the ages of the Mt. Wilson stars, and plotted these new ages in Fig.1 using
large and small squares for measured and calculated periods. 
Comparing with the $+$ signs used earlier for the SDJ91 
relationship, the reader can see at a glance how the ages of young stars are 
correspondingly reduced, and how the ages of the old stars remain unchanged.
For subsequent figures, we will use the
Donahue (1998) relationship instead of that from SDJ91.

Fig.2, which compares the isochrone and activity ages for those PH stars 
where both are available suggests that systematic errors still remain in 
assigning chromospheric ages. Shortward of about 8 Gyr isochrone age, the 
chromospheric ages are clearly higher by about 0.1 dex and longward of this, 
the sense of the disagreement is reversed and the difference approaches 
0.3 dex. Almost all of the latter stars are in the $1 M_{\odot}$ bin.
Open cluster data presented later suggest a downward revision of
F star activity ages. Interestingly, as we shall see, rotational evidence
also suggests that the activity ages of earlier-type stars should be 
reduced considerably while those of later types could be somewhat increased.

In this connection, we also would like to point out that neither the
Soderblom, Duncan \& Johnson (1991) relationship nor the  Donahue (1998) 
relationship consider the
contribution from rotation to $R'_{HK}$, which we know must exist from work
by Noyes et al. (1984), which itself does not consider age variation.
Chromospheric activity is known to be a 
function both of rotation (Noyes et al. 1984) and age (Wilson 1963), and 
probably spectral type, included in the work of Noyes et al. (1984) via the 
convective overturn timescale, which is necessary to calculate the Rossby 
number.
Also, all the stars in Soderblom et al. (1991), regardless of mass, have been 
lumped together into one bin. This might be relevant in understanding why the 
ages seem to be exaggerated for the $1.2 M_{\odot}$ stars.
In this context, it is worth noting that
Suchkov and Schultz (2001) argue for younger ages, `not very different from
the Hyades, ~700 Myr' for the planet-bearing F stars, based on various
considerations, especially variability in the Hipparcos $H_p$ band but it is
unclear to the author how this new age indicator can be related to other,
more road-tested, ones.

On a more positive note, though, systematic errors in the activity ages
for the PH stars would be mirrored in those of the Mt. Wilson stars and 
rotational comparisons would remain valid. 
This fact, combined with the discussion above, suggests that once the ages
are taken into account, there is no 
rotational difference between the Mount Wilson and PH star data, and the
present indications are that the PH star periods are entirely consistent with 
the Mount Wilson periods, provided the activity ages of stars in the 
$1.2 M_{\odot}$ bin are revised downward by 0.2 or 0.3 dex.

	\section{Open Cluster Data}

The rotation of open cluster stars is a large- and complex-enough issue to 
be considered on its own. However, since it has been discussed
in great detail elsewhere (eg. Barnes \& Sofia 1996; Krishnamurthi et al. 
1997; Sills et al. 2000; Barnes et al. 2001), 
we will only mention it briefly. 
Determination of the rotation rates of open cluster stars has become an
industry in its own right and has largely guided the
construction of theoretical stellar models. This has occured mostly because 
of the large sample sizes available, at least in principle, and because the
ages of open cluster systems can be derived meaningfully, permitting a
comparison to theory.
However, most of these measurements pertain to very young systems and thus 
there is as yet no leverage to constrain models of older stars.

	\subsection{Comparison of OC data with PH stars }

The rotational data from several open clusters have been added schematically
(displayed using dark grey boxes) in the upper panels of Fig.3 which also 
display the previously discussed PH star data. 
NGC 3532 and the Hyades are marked individually while the box representing 
the young clusters includes data for IC 2391, IC 2602, Alpha Per and the 
Pleiades. A review of cluster rotation period data may be found in 
Barnes (2000). 
These panels show that the PH star data are in reasonable agreement with
the open cluster data, to within the age uncertainties, in the sense that one 
can imagine almost all the PH star periods displayed evolving out of younger
open cluster-type periods. (This assumes, of course, that open cluster stars
and field stars have similar rotational properties.) Given the age 
uncertainties, we cannot at the present time, say anything more definitive.  

	\subsection{Comparison of OC data with Mount Wilson }

In the lower panels of Fig.3, we now compare the open cluster data to the
Mt. Wilson stars, assigned ages according to Donahue (1998). 
The $0.8 M_{\odot}$ and $1.0 M_{\odot}$ panels show that these Mt. Wilson stars 
are reasonably consistent with the open cluster data currently available. 
The same cannot be said of the stars in the $1.2 M_{\odot}$ panel,
where the open cluster and Mt. Wilson stars seem to diverge significantly.
The open cluster periods are very unlikely to be biased against fast rotators
and neither do their ages seem adjustable by large-enough factors. On the
other hand, most of the Mt. Wilson stars have measured (rather than calculated)
periods and the ages of the younger ones have been already corrected by the 
Donahue (1998) relationship to make them consistent with open cluster ages.
It is unclear to the author how to solve this problem except to suggest
speculatively that the Donahue (1998) and SDJ91 formulae overestimate the ages
of the early-type stars. A correction along these lines would also
alleviate the problem of 10 Gyr ages for some F-type stars.

Young open clusters obviously contain more rapid rotators than do the young
Mount Wilson or PH stars. These are the so-called ultra-fast rotators, or UFRs
(Van Leeuven \& Alphenaar, 1982; Van Leeuwen, Alphenaar \& Meys, 1987). More 
about these presently. The implication 
for the Mount Wilson and PH stars is simply that they include very few
stars younger than a few hundred million years.

	\subsection{The Skumanich relationship }

The lower panels of Fig. 3 also display lines representing 
the Skumanich (1972) relationship, $v \propto t^{-1/2}$, for initial periods 
at 100 Myr of 1, 2, 4 \& 8d. These are not intended to be fits to the data
but merely to suggest how a star starting out with a certain period at 100 Myr
would evolve, were it to obey the Skumanich relationship exactly. 
It is important to emphasize that this relationship is based not on
any model but on Skumanich's (1972) fit to observations of cluster stars.
This initial fit binned together stars of varied mass in each cluster, and 
merely used the $v \sin i$ information available then. The plots here 
demonstrate that such a fit is still approximately valid for main-sequence 
stars, when binned by mass, and in the absence of the $\sin i$ ambiguity, 
with its attendant implications for stellar magnetic fields (cf. Kawaler 1988, 
Barnes \& Sofia 1996). These considerations suggest, furthermore, that PH 
stars are subject to the same angular momentum loss mechanism that 
characterizes `normal' main-sequence stars. 

The ultra-fast rotators, as has been noted earlier (eg. Stauffer 1994),
cannot be explained within the same framework but as shown in Barnes \& Sofia
(1996), can be
explained by including pre-main sequence evolution and magnetic saturation,
which suggests a reduced angular momentum loss rate for fast rotators
(cf. MacGregor and Brenner 1991). 
Also, the TTauri stars (ages $\simeq$ 1 Myr), not plotted here, are slow
rotators, at odds with the Skumanich relationship, and to understand how their
rotation rates could lead to the measured rates of older stars, detailed 
stellar models are needed. We now turn to a discussion of these models in 
the following section, and others that include disk-related stellar spindown.

To conclude this section, we merely note that Fig.3 demonstrates that although 
the overlap on the time-axis between the Open Cluster star data and the PH star
data is minimal, there is no evidence yet to suggest major disagreement.
The outliers HD 95128, HD 117176, and HD 177830 appear to be evolved stars and 
will be discussed in section 6.
 
There appears to be an 
inconsistency in the periods or ages or both of the early-type open cluster 
and Mt. Wilson stars. In passing, we note that, ignoring
the UFRs and perhaps independently, these basic Skumanich-type overlays on the 
data suggest shorter starting periods for higher-mass stars, proposing a
differentiation in the initial conditions that had hitherto been considered 
unwarranted.
Finally, we note that in the absence of detailed stellar models, there is 
no reason to suspect that the PH star data would have looked any different from
the OC star data when the PH stars were younger.

	\section{Rotating evolutionary stellar models}

A large body of work, both observational and theoretical, over decades, has
resulted in the formulation of a fairly general picture of the rotational 
evolution of solar-type stars. This includes internal angular momentum 
transport, wind-induced angular momentum loss on the main sequence leading to
a Skumanich-type slowdown for older stars, magnetic saturation, pre-main 
sequence evolution and now, disk-interaction in the earliest phases
(Kraft 1967; Kippenhahn \& Thomas 1970, Endal \& Sofia 1976, 1978, 1979; 
Van Leeuwen \& Alphenaar 1982; Radick et al. 1987; Kawaler 1988;
Pinsonneault et al. 1989; MacGregor \& Brenner 1991; Charbonneau \& MacGregor 
1992; Edwards et al. 1993; Stauffer 1994; Chaboyer, Demarque \& Pinsonneault 
1995; Cameron, Campbell \& Quaintrell 1995; Bouvier, Forestini \& Allain 1997; 
Queloz et al. 1998; Sills et al. 2000; etc.).

A series of modern models of the rotational evolution of late-type stars are 
available (eg. Barnes \& Sofia 1996; Sills, Pinsonneault \& Terndrup 2000; 
Barnes, Sofia \&
Pinsonneault 2001). Such models give the rotation rate, under varying 
circumstances, as functions of the age of the star. Placing stars of known 
ages and periods on these plots enables us to discover anomalies, if any, and
under certain conditions, their probable rotational histories.

Details about these models, generated using the Yale Rotating Evolutionary
Code (YREC), have already been published elsewhere 
(Pinsonneault et al. 1989; Chaboyer, Demarque \& Pinsonneault 1995; Barnes \&
Sofia 1996; Barnes, Sofia \& Pinsonneault 2001). 
In this instance, it suffices to say that, in addition to the standard stellar 
evolutionary calculations, they model rotating stars as a series of nested, 
deformed shells, 
with internal angular momentum transport via instabilities for 
differentially-rotating models and a parameterized wind that drains angular
momentum from the surfaces of stars with outer convection zones.
By including pre-main sequence evolution and invoking appropriate amounts of 
magnetic saturation and disk-interaction, they are generally able to evolve 
the fast- and slow-rotating TTauri stars respectively into the ultra-fast 
rotators and slow rotators in young star clusters 
The fastest models at all ages are those that are started off at the birthline
with 4d periods and no disk-interaction in the TTauri phase (prior to the
stellar ages plotted in this paper), while the slowest ones are those started 
off with 16d periods and one-to-several Myr of disk-locking.
Intermediate cases lie between these curves (see Fig.4).
These models have been calibrated in such a way that a disk-free solar model 
with a starting period of 10d and no magnetic saturation reaches the solar 
rotation rate at solar age. This allows a limited amount of variation in the 
model rotation rates at large ages\footnote{One could argue that the Sun should be modelled among the disked systems but the historical development of the field has thus far resisted thinking of the Sun as special, preferring (justifiably) to use the Occam's Razor approach of evolving the average (non-disked) rotator into the solar period at solar age. This might have to be reconsidered at some point.}.
Finally, we note that the magnetic field configuration has been chosen to 
match the Solar case\footnote{The Mt. Wilson data also suggest that the magnetic fields and cycles of G \& K stars are similar to that of the Sun.} (as 
discussed in Barnes \& Sofia (1996)) but nature might be considerably less 
restrictive, 
allowing more dipole-like configurations, which would result in somewhat 
faster rotators at later ages. However, note that the general validity of
this configuration in describing Mt. Wilson stars, PH stars and all but the
UFRs among open cluster stars and TTauri stars has been demonstrated in Fig.3,
in the previous section of this paper.

Fig.4 shows the ranges in rotation period allowed by the models against
stellar age (modulo the calibration and field configuration variations
discussed above). Solid-body (SB) and differentially-rotating (DR) models are
plotted in the upper and lower panels respectively, against stellar mass.
The lowest curves in both cases represent the evolution of the fastest 
rotational models, started off without disk-locking, with 4d periods on the 
stellar birthline. 
The uppermost curves in each frame represent the slowest expected rotators,
started off with 16d periods on the birthline\footnote{They speed up on the pre-main sequence, hence have about 10d periods at the main sequence starting ages for these figures.} and subjected to 3Myr and 20Myr of disk-locking 
respectively for the differentially-rotating and solid-body cases. 
Additionally, for the DR cases, we have displayed disk-free models with 16d 
starting periods (second curve from the bottom), to show the narrow range 
predicted for disk-free models, and models with 16d starting periods and 1 Myr 
disk lifetimes (third curve from the bottom), to show the rotational 
variation that could be generated by the simple disk interaction scenario.
Thus, if the models are appropriate, all observed stars should lie within the 
ranges shown (modulo their age or period uncertainties). Furthermore, if
planetary systems have their origins in circumstellar disks (how else could 
they form?), and disk-interaction results in angular momentum loss, as these 
models assume, the planet host stars should appear above the disk-free model
predictions, among the slower rotators. Thus, data points above the the solid 
lines would further the evidence for disk-locking but finding significantly 
faster rotators than the fastest among the models\footnote{There is some evidence in open clusters that the fastest rotating models should be a bit faster to account for the observations, but this does not change the conclusions of this paper} would be problematical. 

The issue of the time-scale of redistribution of angular momentum within stars 
has not been resolved yet, and consequently we plot both solid-body and 
differentially-rotating models, representing the range of possibilities. 
The issues of disk-locking itself, and its rotational consequences are being 
actively debated (Stassun et al. 1999; Herbst et al. 2000). At the present 
time, it appears that solar-type stars display some evidence of being braked 
by disks, but the evidence is considerably more ambiguous for lower mass 
stars. 
Finally, we note that the models were generated assuming the simplest, uniform,
starting conditions for stars of varying mass. The data now seem to suggest 
otherwise, and future grids of models might have to relax this assumption.

	\section{Comparison between models and planet host stars}

Fig.4 compares the host star data and the models.
The open cluster and Mount Wilson data discussed previously are also plotted
for comparison to the models,
which illustrate the range of rotational behavior currently 
producible, and permissible, as detailed in the previous section, for 
both solid-body and differentially-rotating stellar models (upper and lower
panels respectively).
We note here that because of the inclusion of pre-main sequence evolution and
magnetic saturation, the models can be coaxed into reproducing
the UFRs in young star clusters, although they might not be going far enough
for the more massive stars, partially because of the choice of uniform starting
periods for all masses. Models that relax this asumption will doubtlessly 
fit the open cluster data better. Nevertheless some conclusions may still be 
drawn.

The PH star data from Table 1, binned by stellar mass, are overplotted in the 
same figure as before, the larger and smaller asterisks indicating, 
respectively, measured and calculated rotation periods. 
The largest square represents the Sun and the higher-mass stars listed in 
Table 2 (but not modelled here) are also indicated in the $1.2 M_{\odot}$ 
panel using large and small `x'es for measured and calculated periods.

	\subsection{ $0.8 M_{\odot}$ models}

For the $0.8 M_{\odot}$ case, there is reasonable agreement between the data 
and the models, given the number of host stars available in this mass range. 
Most of these are calculated periods (see Table 1) but they agree with those 
of the Mount Wilson sample.
Given the data error bars and modulo some tuning of the models, either the
SB or the DR models would work, although the discovery of more systems with
a wider range of rotation rates would tend to favor the DR models. As of now,
it is too early to tell. However, the Mount Wilson data themselves suggest a 
spread in rotation rates that is absent in the solid-body models. Finally, a 
comparison of the PH stars only with the DR models does not suggest any 
necessity for disk-regulated rotation. The agreement would be even better,
were the ages of the lower-mass Mt. Wilson stars to be increased by a tenth
of a dex or two, as our overall conclusions suggest.

	\subsection{$1.0 M_{\odot}$ models}

The situation with the $1.0 M_{\odot}$ models, for which we have more 
comparison data points, is more complicated. The Solar rotation rate is matched
by both the DR and SB models, as it must by definition, since the overall 
angular momentum loss rate is adjusted so that a suitable stellar model
matches the solar rotation rate at solar age (cf. Barnes \& Sofia 1996; 
Barnes, Sofia \& Pinsonneault 2001).
The other stars have a fairly broad distribution of rotation rates, some of it
doubtlessly arising from the age uncertainties of the individual stars. 
Two stars are rotating considerably slower than the others and one somewhat 
faster than expected for its age.

The slow rotator HD 95128 (P=74d; uncertain) is classified as G1V but it 
is unclear that this star is even on the main sequence (see Henry et al. 2000
and references therein). Given the uncertainty in both its rotation period and 
evolutionary state, it is unclear that any further discussion of it is 
warranted. The other slow rotator is HD 177830, but this star has a calculated,
rather than measured, period. Furthermore, the long period can be attributed
to the fact that it appears to have evolved up the giant branch 
(Gonzalez et al. 2001).

HD 143761 seems to be spinning abnormally fast for its age, but we defer a
discussion of it to section 7, which deals with tidal spin-up.
The other fast rotators are still somewhat problematical for the models,
as the $1.0 M_{\odot}$ panels plainly show.
There are enough of these stars with measured periods that it is clear that
the PH star data cannot be blamed. Furthermore, they agree with the Mt. Wilson
stars. It seems inescapable here that the PH star and Mt. Wilson datasets
have a greater rotational dispersion than that generated by the models. 
In view of the discussion of the $1.2 M_{\odot}$ models below, it would appear 
that the models lose angular momentum at a higher rate than the data demand, 
or else the assumptions made for the initial conditions are incorrect. 
There isn't any major disagreement between the open cluster and Mount Wilson 
data for these stars, provided that the forumula from Donahue (1998) is used 
to derive the ages, although a reduction of the Donahue ages by a tenth of
a dex or two would result in a far better match to the open cluster data.

	\subsection{$1.2 M_{\odot}$ models}

The situation is the worst for the $1.2 M_{\odot}$ models, which are clearly
rotating too slow to explain the open clusters, the PH stars, and the Mt. 
Wilson stars, probably because of the over-simplified starting conditions
mentioned earlier. Of the four PH stars in this bin, one has an uncertain
period and the remaining three are calculated, so it is not clear that any
strong conclusion may be drawn, although it is true that they line up well
with the open cluster data. We have plotted them here using the isochrone
ages, which are smaller than the activity ages, and explain at least some
of the offset between them and the Mt. Wilson stars, which seem to demand
considerable age reduction. Finally, we note that the slowest of them, 
HD 117176 = 70 Vir appears to have evolved off the main sequence 
(Gonzalez 1998).

	\subsection{Higher-mass PH stars}

Although higher-mass stars are not modeled in this context because of 
differences in their rotational, magnetic and structural properties, we
have plotted them also in Fig. 4 to show the essential continuity in their 
rotational
characteristics, with those of the lower-mass stars. They populate a zone
with high rotation rates (see Table 2), explicable in terms of the break in 
the Kraft (1967) curve. These higher-mass stars do not have surface 
convection zones and do not suffer the wind-induced angular momentum 
loss associated with solar-type stars. Hence, they are all very fast rotators.
The one slow rotator among them, HD 38529, has a calculated period, and is 
located off the main sequence (Gonzalez et al. 2001), removing it effectively
from rotational considerations. Thus,
the PH star, Mt. Wilson, and open cluster data all suggest that the models
need to be even more aggressive in producing fast rotators. However, in at
least one case, the models cannot be the only problem and 
$\tau Boo = HD 120136$ must be excluded from normal rotational considerations
because it is spinning so fast that it raises the issue of tidal spin-up.

		\section{Tidal spin-up}

Even by the standards of the higher-mass stars, $\tau Boo$ is spinning 
extremely fast, and the measured rotation period of 3.2d (Henry et al. 2000)
is almost the same as the orbital period of 3.312d (Butler et al. 1997).
The low eccentricity and other orbital parameters suggest circularization.
Furthermore, the very short spin-up timescale of 0.8 Gyr (Trilling 2000), 
less than the 1.5-2 Gyr age of the system, suggests that spin-up has indeed
occured. In other cases, it is not yet possible to say definitively that
spin-up has occured. Of all the PH star rotation periods presented thus far,
only five others have been securely measured. Of these, $\rho Cr B = HD 143761$
with a measured period of 19d stands out because of its 11 Gyr isochrone age.
However, using the activity age of 6-6.5 Gyr would make the rotation rate
appear normal, and in any case, the spin-up timescale, calculated using 
equation (4) from Trilling (2000) is greater by orders of magnitude. 
HD 209458 has a spin-up timescale of 6.5 Gyr, comparable to its age of 
3-4 Gyr, but no measured rotation period. The calculated period of 15.7d is
much larger than the 3.5d orbital period of the planet.

Approaching this issue from a different, radial velocity perspective, we 
identify ten systems with short orbital periods, almost circular orbits, 
and some rotational and age information. Because the orbital information
suggests circularization, we have examined the (admittedly meager) rotation
information for spin-up, and the result is that these systems are scattered
widely and no trend is visible, except as noted above. This difference 
between the orbital and rotational results is actually to be expected because
a planet's circularization timescale is far smaller than the star's spin-up
timescale (cf. Goldreich and Soter 1966; Trilling 2000, especially Table 2). 
$\tau Boo$ is evidently special.

		\section{Conclusions} 

The heterogeneous nature of the inputs to this investigation make a synthesis
difficult to achieve, but they also reveal inconsistencies. These are 
numerous enough that it is not yet possible to suggest that PH stars have 
abnormal rotation rates. 
The principal result of this work is that it is, or soon might be, possible
to consider the rotation of all solar-type stars, whether near or far, young
or old, planet host or not, in a single, unified scheme that is amenable to
modeling.

The scarcity of measured rotation periods for PH stars is a hindrance to
progress, and it is clear that their ages, and those of the Mt. Wilson stars,
can be improved. These uncertainties evidently inflate the rotation 
distribution relative to that in open clusters. The overall evidence suggests
that activity ages for early G and late F stars are too long by a few tenths
of a dex. Those for K dwarfs might correspondingly be too short by a tenth of 
a dex or two.

Selection of candidate PH stars for radial velocity studies based on low
rotation and/or activity has apparently resulted in a PH star sample biased
towards older stars relative to nearby field dwarfs eg. the Mt. Wilson sample.
The presently available rotational models display inadequacies, most notably
in producing fast-enough G-F stars. It seems premature to attempt to choose
between different rotational models based on these inputs.

Finally, the difference between the number of systems displaying orbital
circularization ($\sim 10$) and the number suggesting tidal spin-up 
($\sim 1$) is explicable in terms of the difference in time-scales for
the two effects.

We began this investigation hoping to find a rotational signature of 
star-disk interaction, either extant or in the past for planetary systems,
that potentially could have been useful in discovering new systems. 
We have found, instead, that there is no rotational difference between
the host stars of extra-solar planets and `normal' solar-type stars. 
Nor do the present data suggest a variation from open cluster stars. 
However, the Mt. Wilson stars, which are themselves consistent with the
PH stars, display inconsistencies with open cluster stars.
In our opinion, these inconsistencies are not fatal, and do not
threaten the broader conclusion that PH stars are normal.

This work may consequently be interpreted as another piece of evidence
that stars with detected extra-solar planets are not 
especially different from other main-sequence stars. 
It suggests that circumstellar matter and/or planets are probably ubiquitous,
reflecting angular momentum considerations. Condensation of a star from a 
cloud would violate these considerations unless 
some material is left behind with large specific angular momentum. This is
no different from the situation in the solar system.
This conclusion is not new, for astronomy has steadily disproven a series of
notions of our uniqueness. 
Circumstellar material is probably the norm, rather than the exception.

	\section{Speculative and broader issues: }

In a certain sense, stars with extra-solar giant planets are binary stars.
Patience and Ducheme (2000) have presented some preliminary evidence that
binary stars with close companions seem to be rotating faster than those
with distant ones. If verified, this would be very interesting. 
Another interesting, if speculative, connection between normal solar-type 
stars and planets is 
the suggestion by Rubenstein and Schaefer (2000) that superflares on ordinary
solar-type stars are caused by magnetic reconnection events between the
star and a nearby planet. Is seems that this sort of phenomenon would work
to slow down the parent star, though, so it is unclear that this can be
reconciled with the Patience and Ducheme (2000) observation or with the
Mt. Wilson star rotation rates.
Quite apart from these speculations, the fact that planet host stars are 
similar to nearby late-type stars ought to be combined with other available 
information to place it in a broader context.

Although the problems presented in this paper are real enough, they are not 
fatal and are probably even quickly solvable, so we should not allow them to
obscure the bigger picture, where PH stars are the norm, rather than the
exception. In this picture, the angular momentum of the parent cloud that
does not appear as orbital angular momentum in a multiple star system must
lurk in the vicinity of the individual star.
Both disk and star are formed together; different parts of essentially the
same object. There cannot be one and not the other. 
This might be related to the agreement between the meteoritic, the 
zircon-based earth age, and the astrophysical solar age. 
Such a picture might not preclude a fountain or a bipolar-type outflow
in the beginning.

A planetary system is formed in time. 
The radial distribution of matter is not presently known and may vary. 
However, this scenario suggests that planetary systems ought to be ubiquitous 
among solar-type stars. This creation leaves an Oort-type cloud behind if a 
massive object is formed, and a Kuiper-type belt could blend the distinction 
between the inner and outer objects. This could explain the similarities 
between properties of PH stars and otherwise `normal' stars.

The formation of the disk solves the angular momentum problem in star 
formation. Almost all of the angular momentum in the system resides in the 
disk. The little angular momentum in the central star (which happens to be
easier to measure) is gradually transported outwards, as it must, because
stars tend to evolve in the direction of increasing central concentration
during their lifetimes. This process is fastest on the pre-main sequence,
suggesting efficient angular momentum transport during this phase, perhaps
through star-disk interaction. 

This star-disk interaction does not have to be an all-or-nothing mechanism, 
as suggested in disk-locking scenarios, but could be gentler and more gradual, 
especially as star-disk interaction may persist in the form of star-planet 
interaction if appropriate conditions obtain. Star-planet interactions, 
magnetic or otherwise, may have observable consequences, as for instance, 
flaring during episodes of magnetic reconnection, as suggested by some.

In general, during the life of the star, an almost ineluctable process of 
centrally concentration, appropriate angular momentum transport mechanisms 
will arise to undertake this transport, 
internally via instabilities, externally via winds, or superficially in the 
early stages via star-disk interaction. Otherwise, concentration cannot occur.
Angular momentum considerations demand that some material remain outside
while some goes inside. 

Finally we note that the striking similarities and differences between
young stellar/planetary systems, where transport can occur, and spiral 
galaxies, where transport is difficult, bear some rumination.

{\it Acknowledgements.} 
SB would like to acknowledge the McKinney Foundation and the NSF for support 
under AST-9731302 and AST-9986962 at the Univ. of Wisconsin. Some of these 
results were obtained as part of SB's PhD dissertation work at Yale 
University, which supported him for several years through a student fellowship.
SB would like to thank Sabatino Sofia, Bill Cochran and the referee for their 
helpful suggestions.
This research has also made use of the SIMBAD database, 
operated at CDS, Strasbourg, France and the Extrasolar Planets Encyclopaedia 
at http://www.obspm.fr/planets maintained by J. Schneider.


\begin{table}
\caption{Characteristics of solar-type stars with planetary companions} \label{tbl-1}
\begin{center}\tiny
\begin{tabular}{lrllclcllccc}
\tableline
(1)&(2)&(3)&(4)&(5)&(6)&(7)&(8)&(9)&(10)&(11)&(12)\\
Name&HD&SpType&B-V&$T_{eff}$&$logR'_{HK}$&$P_{meas}$(d)&$P_{calc}$(d)&$M/M_{\odot}$&IsoAge(Gyr)&ActAge(Gyr)&Sources\tablenotemark{a}\\
\tableline
\\

...		  &192263 &  K2V & 0.938 & 4840	&-4.37  &  .... &  9.5 & 0.75 & ...  & 0.3 &	20,11,08,20,--,20,20,--,31 \tablenotemark{b}\\
$\epsilon$Eri     & 22049 &  K2V & 0.881 & 5180	&-4.455 &  11.7 & .... & 0.78 & $<1$ & 0.7 &	11,11,06,02,05,--,23,23,98\\	
GJ86		  & 13445 &  K1V & 0.812 & 5280	&-4.74	&  .... & 31   & 0.78 & ...  & 2.1 &	12,11,08,32,--,25,25,--,98 \tablenotemark{c}\\
...		  &130322 &  K0V & 0.781 & 5340	&-4.39	&  .... &  8.7 & 0.79 & ...  & 0.3 &	09,11,08,31,--,26,26,--,31\\
...		  &168443 &  G8IV& 0.724 & 5430	&-5.08	&  .... & 37   & 0.84 & 10.5 & 7.4 &	15,11,15,15,--,15,15,31,31 \tablenotemark{d}\\
GJ3021		  &  1237 &  G6V & 0.749 & 5540	&-4.44	&  .... & 10.4 & 0.9  & ...  & 0.6 &	11,11,21,32,--,17,17,--,17\\
...		  &210277 &  G7V & 0.739 & 5540	&-5.06	&  .... & 40.8 & 0.92 & 12   & 6.9 &	15,11,09,15,--,09,09,09,15\\
\\
\\
...	  	  &222582 &  G5	 & 0.648 & 5735	&-5.00	&  ....	& 25   & 0.95 & 11   & 5.6 &	11,11,31,31,--,99,31,31,31\\
...		  &217107 &  G7V & 0.744 & 5560	&-5.00	&  .... & 39   & 0.96 & 12   & 5.6 &	07,11,08,07,--,07,07,31,31\\
$\rho$CrB	  &143761 &  G2V & 0.601 & 5790	&-5.048	&  19   & 19.9 & 0.96 & 11   & 6.6 &	11,11,08,10,10,10,10,34,98 \tablenotemark{e}\\
16CygB		  &186427 &  G5V & 0.661 & 5740	&-5.115 &  31:  & 27.4 & 0.97 &  9   & 8.3 &	11,11,08,10,10,10,10,34,98\\
...		  &195019 &G3IV-V& 0.662 & 5690	&-4.85	&  .... & 22   & 0.98 &  ... & 3.2 &	11,11,08,07,--,07,07,--,07\\
Sun		  &...... &  G2V & 0.64  & 5777	&-4.89	&  26.1 & .... & 1.00 & 4.56 & 3.7 &	04,04,04,32,05,--,--,24,98\\
...		  &187123 &  G3V & 0.646 & 5820	&-4.93	&  .... & 30   & 1.0  & 4    & 4.3 &	25,11,08,03,--,03,03,09,98 \tablenotemark{f}\\
...		  &  6434 &  G3IV& 0.613 & 5845	&-4.89	&  .... & 18.5 & 1.0  & ...  & 3.7  &	19,19,19,32,--,19,19,--,19 \tablenotemark{g}\\
...		  &121504 &  G2V & 0.593 & 6080	&-4.73	&  .... & 14.8 & 1.00 & ...  & 2.8 &	11,11,21,32,--,19,19,--,19 \tablenotemark{h}\\
...	   	  & 12661 &  K0  & 0.71  & 5714	&-5.12	&  ....	& 36   & 1.01 & 8    & 8.4 &	11,11,31,31,--,99,31,31,31\\
23Lib	  	  &134987 &  G5V & 0.691 & 5715	&-5.01	&  ....	& 30.5 & 1.02 & 9    & 5.8 &	11,11,31,31,--,99,31,31,31\\ 
...		  &177830 &  K0IV& 1.062 & 4818	&-5.28	&  ....	& 65   & 1.03 & 11   &13.5 &	11,27,31,31,--,99,31,31,31\\
47UMa		  & 95128 &  G1V & 0.617 & 5850	&-5.041	&  74:  & 21.0 & 1.03 & 6.3  & 6.5 &	22,22,08,10,10,10,10,35,98 \tablenotemark{i}\\
55Cnc		  & 75732 &G8V-K0IV&0.86 & 5250	&-4.949	&  39   & 42.2 & 1.05 & 3.6  & 5   &	01,01,33,10,10,10,33,33,01 \tablenotemark{j}\\
51Peg		  &217014 & G2.5V& 0.67  & 5760 &-5.068	&  21.9:& 29.5 & 1.05 & 5.1  & 7.1 &	04,04,08,10,10,10,10,35,98 \tablenotemark{k}\\
... 		  & 92788 &   G5 & 0.694 & 5775	&-5.04	&  ....	& 32   & 1.05 & 4.2  & 6.4 &	11,11,31,31,--,99,31,31,31\\
79Cet		  & 16141 &  G5IV& 0.67  & 5777	&-5.05	&  ....	& 29   & 1.05 & 8.5  & 6.7 &	11,11,31,31,--,99,31,31,31\\
...		  & 82943 &   G0 & 0.623 & 6010	&-4.95	&  .... & 20.9 & 1.05 & ...  & 5   &	22,11,21,14,--,14,14,--,21 \tablenotemark{l}\\
...		  & 52265 &  G0V & 0.572 & 6060	&-4.91	&  .... & 14.6 & 1.05 & 2.1  & 4   &	11,11,21,14,--,14,14,31,31 \tablenotemark{m}\\
...		  &108147 &F8/G0V& 0.537 & 6260 &-4.78  &  .... &  8.7 & 1.05 &  ... & 2.5 &  	11,11,21,32,--,14,14,--,98 \tablenotemark{n}\\
\\
\\ 
70Vir		  &117176 &  G5V & 0.714 & 5530 &-5.115 &  31:  & 35.8 & 1.10 & 7.7  & 8   &	11,11,08,10,10,10,10,35,10 \tablenotemark{o}\\
...		  &209458 &  G0V & 0.574 & 6000 &-4.93  &  .... & 15.7 & 1.1  & 3    & 4.3 & 	22,16,16,31,--,14,16,31,98\\
...		  & 75289 &  G0V & 0.578 & 6120 &-5.00  &  .... & 16   & 1.15 & 4.5: & 5.6 &	12,11,08,32,--,26,26,26,26 \tablenotemark{p}\\
$\iota$Hor 	  & 17051 &  G0V & 0.561 & 6100 &-4.65  &  .... &  8.3 & 1.19 &  1   & 1.6 &	13,11,13,32,--,13,31,31,98 \tablenotemark{q}

\end{tabular}
\end{center}

\tablenotetext{a}{The 9 sources correspond respectively to the 9 columns (3) through (11) where the sources are - 
01: Baliunas et al. 1997; 
02: Baliunas, Sokoloff \& Soon 1996;  
03: Butler et al. 1998; 
04: Cayrel de Strobel 1996; 
05: Donahue et al. 1996; 
06: Drake \& Smith 1993; 
07: Fischer et al. 1999; 
08: Gimenez 2000; 
09: Gonzalez et al. 1999; 
10: Henry et al. 2000; 
11: Hipparcos (Perryman et al. 1997);  
12: Houk 1978; 
13: Kurster et al. 2000; 
14: Geneva Obs. Web site at http://obswww.unige.ch/;  
15: Marcy et al. 1999; 
16: Mazeh et al. 2000; 
17: Naef et al. 2000; 
18: Queloz et al. 2000a; 
19: Queloz et al. 2000b;
20: Santos et al. 2000; 
21: Santos, Israelian \& Mayor 2000; 
22: Simbad;  
23: Soderblom \& Dappen 1989; 
24: Tilton 1988; 
25: Trilling 2000;  
26: Udry et al. 2000;  
27: Vogt et al. 2000; 
31: Gonzalez et al. 2001; 
32: Henry et al. 1996; 
33: Gonzalez \& Vanture 1998; 
34: Gonzalez 1998; 
35: Lachaume et al. 1999; 
98: Calculated using eqn (2), taken from Donahue 1998; 
99: Calculated using eqns 3 and 4 from Noyes et al. 1984}
\tablenotetext{b}{Vogt et al. (2000) see power in a rotation period around 26.8d, uncomfortably close to the Doppler velocity period of 24d. The origin of the 3 Gyr age in (25) is unclear.}
\tablenotetext{c}{Presumably the age in (25) has been calculated from activity, since the star is below the main sequence Queloz et al (2000a).}
\tablenotetext{d}{See the discussion about mass, age etc. in (15) and refs therein. (34) suggests an age $<$ 1Gyr from isochrone fit.}
\tablenotetext{e}{$P_{meas}$ has evidently been updated from (02), which lists 17d. The Gonzalez (1998) age appears very large. Ng and Bertelli (1998) also suggest 12.1 Gyr, $0.93 M_{\odot}$.}
\tablenotetext{f}{(09) suggests 1.08 $M_{\odot}$, but overall characteristics, including rotation, suggest a lower mass.}
\tablenotetext{g}{Preliminary information from (19); no refereed publication available yet.}
\tablenotetext{h}{Preliminary information from (19); no refereed publication available yet.}
\tablenotetext{i}{$P_{calc}$ and age estimate make the possibly incorrect assumption that 47Uma is on the main sequence, a fact which might also explain the weakly measured long rotation period (see 10). Ng and Bertelli (1998) suggest 6.5 Gyr, 1.06 $M_{\odot}$. Edvardsson et al. (1993) suggest 6.9 Gyr from isochrones. (34) suggests 8 Gyr from isochrones.}
\tablenotetext{j}{55Cnc could be a subgiant, not a dwarf (see 01), explaining the long rotation period. See Gonzalez \& Vanture (1998)}
\tablenotetext{k}{$P_{meas}$ has evidently been updated from (02), which lists 37d. (10) suggests 3-7 Gyr from activity; (31) suggests 5.5 Gyr from isochrone fit.}
\tablenotetext{l}{Preliminary data from (14); (21) lists 1.08 $M_{\odot}$.}
\tablenotetext{m}{Preliminary data from (14); (21) lists 1.13 $M_{\odot}$. Butler et al. (2000) derive $R'_{HK}=-4.99$.}
\tablenotetext{n}{Preliminary data from (14); (21) lists 1.15 $M_{\odot}$.}
\tablenotetext{o}{According to (34), 70 Vir is almost certainly a subgiant.}
\tablenotetext{p}{Gonzalez and Laws (2000) suggest a 2.1 Gyr isochrone age.}
\tablenotetext{q}{$P_{calc}$ values vary between 7.9d and 8.6d, while isochrones suggest 7Gyr age (see 13); (35) suggests a 3 Gyr age from an isochrone fit.}

\end{table}

\newpage

\begin{table}
\caption{Characteristics of higher-mass stars with planetary companions} \label{tbl-2}
\begin{center}\tiny
\begin{tabular}{lrllclcllccc}
\tableline
(1)&(2)&(3)&(4)&(5)&(6)&(7)&(8)&(9)&(10)&(11)&(12)\\
Name&HD&SpType&B-V&$T_{eff}$&$logR'_{HK}$&$P_{meas}$(d)&$P_{calc}$(d)&$M/M_{\odot}$&IsoAge(Gyr)&ActAge(Gyr)&Sources\tablenotemark{a}\\
\tableline
\\

$\upsilon$And   &    9826 & F8V & 0.536 & 6170 & -4.927	& 14: & 11.6 & 1.31 & 2.9 & 4.3	& 11,11,08,10,10,10,10,35,98 \tablenotemark{x}\\
...		&   19994 & F8V	& 0.575 & 6160 & -4.84	& ....& 13.7 & 1.35 & 3	  & 3.1	& 11,11,21,19,--,19,19,19,19\\
$\tau$Boo	&  120136 & F7V	& 0.508 & 6380 & -4.733	& 3.2 &  5.1 & 1.36 & 1.4 & 2.1	& 11,11,08,10,10,10,10,35,98 \tablenotemark{y}\\
...		&  169830 & F9V	& 0.517 & 6300 & -4.93	& ....&  9.5 & 1.37 & ... & 4	& 11,11,21,14,--,14,21,--,21\\
...		&   89744 & F7V	& 0.531 & 6166 & -5.12	&  9  & .... & 1.4  & 1.8 & 8.4	& 11,11,30,02,02,--,30,31,31\\
...		&   38529 & G4V & 0.773	& 5646 & -4.89	& ....&	34.5 & 1.49 & 3	  & 3.7 & 11,11,31,31,--,99,31,31,31 \tablenotemark{z}

\end{tabular}
\end{center}


\tablenotetext{a}{The 9 sources correspond respectively to the 9 columns (3) through (11)
where the additional sources are - 
28: Allen's Astrophysical Quantities 2000; 
29: Henry et al. 1997; 
30: Korzennik et al. 2000}
\tablenotetext{x}{Ng and Bertelli (1998) also suggest 2.7 Gyr, $1.28 M_{\odot}$.}
\tablenotetext{y}{Evidently the rotation period has been updated from (02), which lists 4d.}
\tablenotetext{z}{The spectral type, color, $T_{eff}$, and rotation period are all consistent with a lower mass star.}

\end{table}

\clearpage

\large{Figure captions}

\figcaption[f1.ps]{The observational data currently available for PH stars (asterisks) compared with those for the Mt. Wilson stars (grey boxes), binned into two categories, younger and older than 2 Gyr (upper panels) and assigned ages (lower panels) according to Soderblom et al. 1991 ($+$es) and Donahue 1998 (squares). The large square represents the Sun and the crosses the higher-mass stars from Table 2. Large and small symbols represent measured and calculated rotation periods respectively. Stars discussed individually and those with uncertain period measurements (marked by colons in this and subsequent figures) are labeled using their HD numbers in the upper panels.\label{fig1}}

\figcaption[f2.ps]{Comparison of Isochrone and Activity ages for planet host stars.}\label{fig2}

\figcaption[f3.ps]{Open cluster data currently available (dark grey boxes) is compared with PH stars (asterisks) and the Mt. Wilson data (light grey boxes)in the upper panels. The lower panels display Skumanich-type ($v \propto t^{-1/2}$) relationships for periods of 1, 2, 4 and 8d at 100Myr, superimposed on the data. Other symbols as in Fig.1.}\label{fig3}

\figcaption[f4.ps]{Comparison between rotational models and planetary system hosts (asterisks). Solid-body (SB, solid lines) and differentially-rotating models (DR, dashed lines) of 0.8, 1.0 and 1.2 $M_{\odot}$ are plotted respectively in the upper and lower panels. The models for the fast rotators in both cases were started off with 4d initial periods and no disk-interaction while the slow rotator models were started off at the birthline with 16d initial periods and 20Myr or 3Myr of disk-locking respectively for the SB and DR cases. The DR models, additionally, include one (second from bottom) with 16d initial period but no disk and one (third from bottom) with 16d initial period and 1 Myr of disk interaction. Other symbols as in Fig.3.}\label{fig4}

\clearpage

\begin{figure}[1]
\plotone{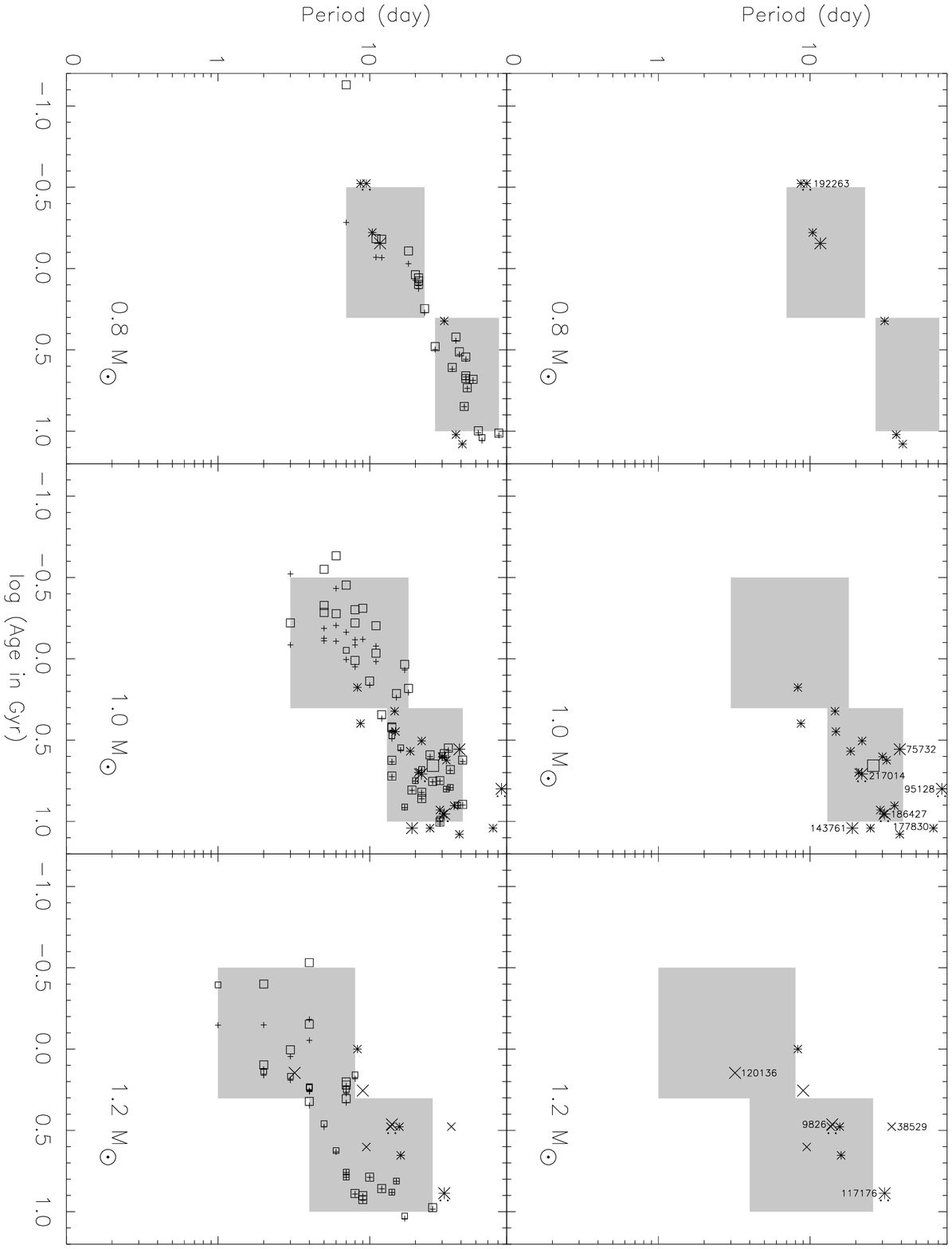}	
\end{figure}

\begin{figure}[2]
\plotone{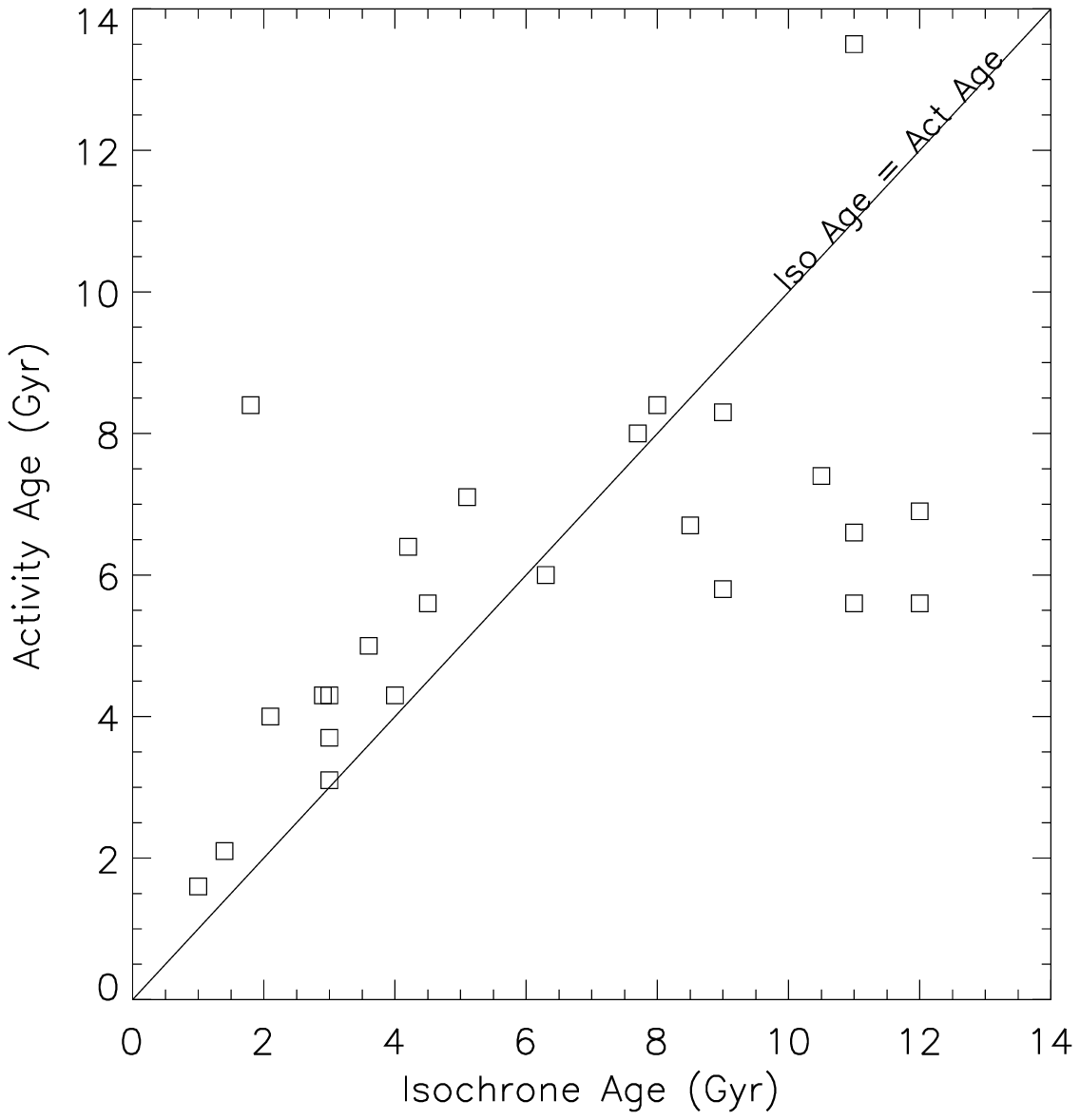}	
\end{figure}

\begin{figure}[3]
\plotone{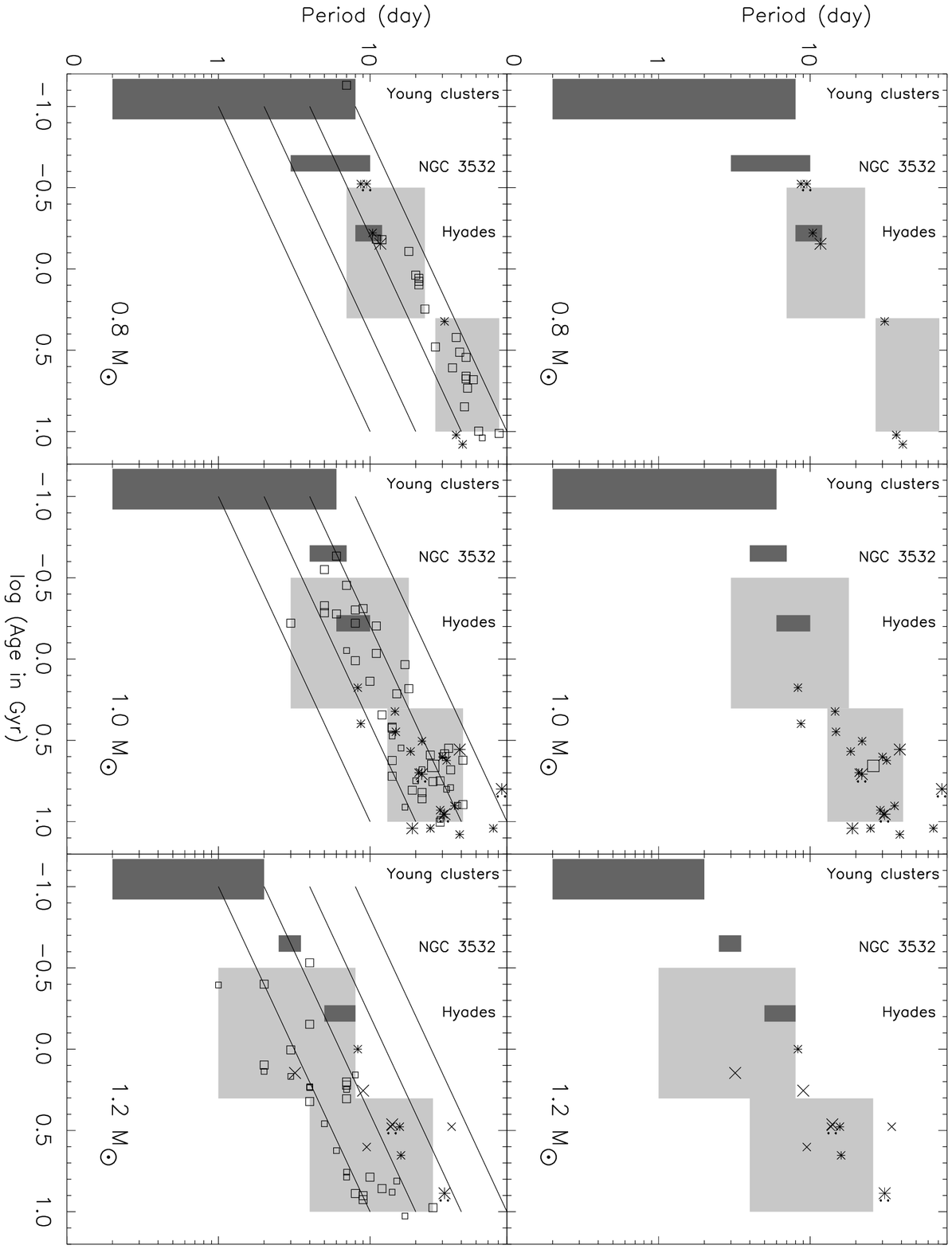}	
\end{figure}
 
\begin{figure}[3]
\plotone{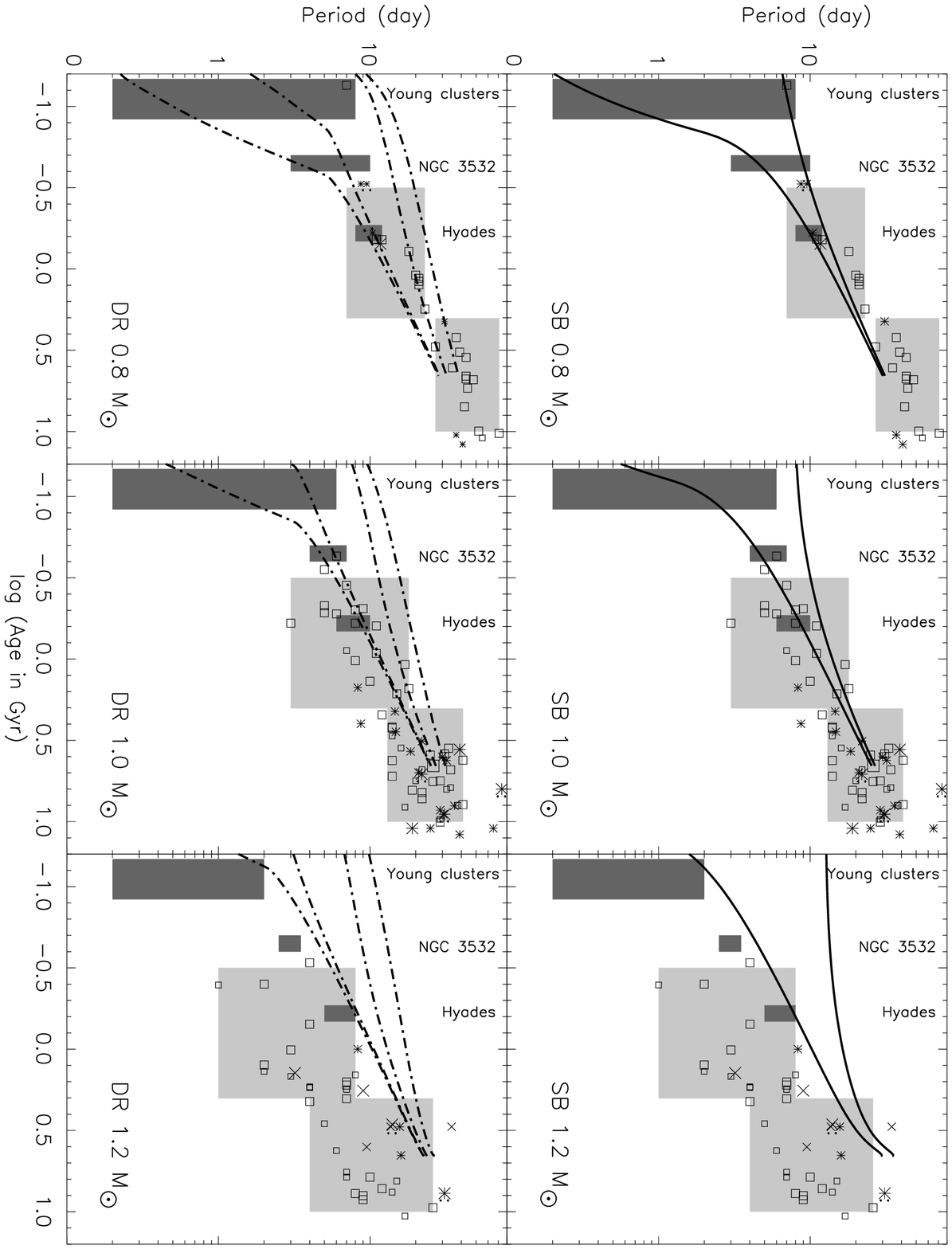}	
\end{figure}

\end{document}